\documentclass{article}
\usepackage{enumitem} 
\setlist[itemize]{itemsep=0pt, topsep=5pt, parsep=0pt, partopsep=0pt}
\RequirePackage[bf,compact,small]{titlesec}
\titleformat*{\section}{\rmfamily\bfseries}
\titleformat*{\subsection}{\rmfamily\bfseries}
\titleformat*{\subsubsection}{\rmfamily\bfseries}
\titleformat*{\paragraph}{\rmfamily\bfseries}
\RequirePackage{microtype}
\usepackage[
  top=1in,
  bottom=1.25in,
  left=1.25in,
  right=1.25in
]{geometry}
\RequirePackage{booktabs}
\usepackage{graphicx}
\usepackage[backend=biber,style=ieee]{biblatex}
\RequirePackage[usenames,dvipsnames,svgnames,table]{xcolor}
\RequirePackage[breaklinks,colorlinks,linkcolor=DarkGreen,citecolor=DarkRed,urlcolor=white!0!blue!99]{hyperref}
\title{Report on Challenges of Practical Reproducibility for Systems and HPC Computer Science}
\begin{document}
\maketitle
\begin{center}
\large
Kate Keahey\\
\textit{Argonne National Laboratory, The University of Chicago}\\[1ex]
Marc Richardson\\
\textit{The University of Chicago}\\[1ex]
Rafael Tolosana Calasanz\\
\textit{University of Zaragoza}\\[1ex]
Sascha Hunold\\
\textit{Vienna University of Technology}\\[1ex]
Jay Lofstead\\
\textit{Sandia National Laboratories}\\[1ex]
Tanu Malik\\
\textit{University of Missouri-Columbia}\\[1ex]
Christian Perez\\
\textit{INRIA}\\[3ex]
\end{center}

\vspace{1em}
\noindent
\textbf{\emph{To cite this version}}:
Kate Keahey, Marc Richardson, Rafael Tolosana Calasanz, Sascha Hunold, Jay Lofstead, Tanu Malik, Christian Perez. Report on Challenges of Practical Reproducibility for Systems and HPC Computer Science. [Report] April 30, 2025. doi: 10.5281/zenodo.15306610.

\thispagestyle{empty}
\newpage
\tableofcontents
\thispagestyle{empty} 
\clearpage             
\pagenumbering{arabic}
\section{Introduction}

There is a broad agreement that in the digital age science should be shared digitally, through artifacts such as code and data, and that adopting practices enabling or facilitating reproducibility of computational results can lead to more robust science and increased scientific productivity \cite{gundersen2018reproducible,krafczyk2021learning,rosendo2020e2clab,stodden2016enhancing}. To support this consensus, the computer science community conducts artifact evaluation (AE) initiatives at conferences and sponsors reproducibility badges awarded by major computing organizations \cite{Boisvert2016Incentivizing,Frery2020Badging}. It is however not clear how much the artifacts, packaged and verified during AE initiatives with so much effort, are used after the one-time verification  as part of individual quest for knowledge. We argue that unless reproducing research becomes as vital and mainstream part of scientific exploration as reading papers is today, reproducibility will be hard to sustain in the long term because the incentives to make research results reproducible won’t outweigh the still considerable costs of reproducibility. Therefore, in addition to seeking ways to ensure that every experiment can be repeated regardless of the effort, we should also explore mechanisms for \emph{practical reproducibility} \cite{keahey2023three}, i.e., a practice where many -- or even most -- experiments are packaged in such a way that they can be repeated cost-effectively. The desire to advance this understanding was the reason for organizing a community workshop on practical reproducibility. 

The objective of the workshop was to \emph{understand and characterize the practical reproducibility problem} -- rather than assess the state of reproducibility solutions or community adoption. With this in mind, the workshop (and the associated Birds-of-a-Feather (BoF) session at SC24) solicited presentations from \emph{reproducibility practitioners}, i.e., experiment \emph{authors, reviewers}, as well as \emph{chairs of reproducibility initiatives} to share experiences, practical insights, and formulate a path forward for community adoption of reproducibility practices. Since many of them participated in AE initiatives and other reproducibility events supported by Chameleon, the workshop also doubled as Chameleon User Meeting \cite{chameleon_user_meetings} even though not every attendee was an active Chameleon user. We also chose to focus on the area of \emph{systems and HPC computer science research} rather than the much broader field of computational science, because it often has unique and particularly challenging requirements (e.g., access to specialized hardware or ability to modify the system on firmware and operating system levels), that often get lost or labeled as ``outliers" when addressed in the broader context. It is our hope, that by addressing the more challenging problem with a narrower scope, we will contribute insight necessary to solve one aspect of the broader problem. Workshop attendance was open to all, and we particularly solicited attendance from open research infrastructure providers as their insight is critical to solving one of the longest-standing and thorny obstacles to reproducibility -- that of resouce availability. 

The resulting report reflects the focus of the workshop: the main two sections, \textbf{Section \ref{challenges}} and \textbf{\ref{recommendations}}, define the challenges of practical reproducibility and present recommendations of the workshop on solutions addressing these challenges. Each of those sections is split into a discussion of the technical challenges/recommendations of packaging or reviewing experiments and the organizational/societal challenges of conducting AE initiatives. The recommendations are labeled by their target audience, i.e., authors or reviewers -- for reproducibility practitioners; organizations -- for conferences; AE initiatives, or other organizations that are, e.g., in a position to influence incentive structures, such as the ACM and IEEE professional organizations); and community -- for members of the community that support the development of the ecosystem of tools and resources for practical reproducibility. The latter contain recommendations for what new tools should be built or improved. Some recommendations contain content relevant to multiple target audiences, e.g., the recommendation to use open, shared infrastructure is potentially relevant to authors, reviewers, as well as AE organizers. \textbf{Sections \ref{organization}} and \textbf{\ref{background}} provide a detailed description of the workshop organization and report writing process as well as define concepts in the context of which the rest of the report is to be interpreted; a reader familiar with the area may want to skip those sections at first and refer back to them as needed. Last but not least, the \textbf{\hyperlink{appendixa}{appendix A}} and \textbf{\hyperlink{appendixb}{B}} contain author/reviewer checklists for artifact preparation which were a significant result of the workshop. We expect that our understanding of what should go in this type of checklist will evolve further; for this reason, the checklists are equipped with instructions on how to contribute insight on an ongoing basis. 

The workshop resulted in much insightful discussion and generated much progress in areas as diverse as recommendations on completeness of artifact description to recommended incentive structures. Some of the most insightful recommendations centered around formulating different methodologies for ascertaining if, and to what extent, an experiment has been reproduced and discussing the balance of cost/incentive, longevity, and findability of artifacts (or the shape of digital libraries that might effectively support them). An unexpected and intriguing recommendation was to use artificial intelligence (AI) for tasks that are notoriously complex in the current reproducibility ecosystem, such as e.g., the establishment of experimental environments or finding a reproducibility condition -- to the point of posing a significant barrier to practical reproducibility. At the same time, skeptics noted that realistic progress would require a large volume of existing and evolving data and thus a large volume of diverse experiments. Lastly, the workshop resulted in practical, detailed, and immediately actionable recommendations for how to package artifacts for practical reproducibility in the form of checklists, which we hope will inform upcoming AE initiatives.

\section{Workshop Organization}
\label{organization}

The Community Workshop on Practical Reproducibility in HPC \cite{practical_repro_hpc2024} convened on November 18, 2024, at Terminus 330 in Atlanta, Georgia. This full-day, in-person workshop was scheduled alongside the SC24 conference (November 17-22, 2024), though not formally associated with the conference in any way. This co-location was deliberate as Chameleon had served as the default reproducibility platform for SC24's AE process \cite{sc24_artifact_evaluation} and had supported over 15 authors -- and many reviewers -- in submitting artifacts to SC24 so that the conference provided a good venue for soliciting their feedback. The workshop was supported by the National Science Foundation (NSF) REPETO project \cite{repeto} (NSF award 2226406).

The workshop's program was structured to facilitate meaningful discussion about challenges in artifact packaging and evaluation. Following an open call for presentations focused on these topics, the planning committee -- comprising Rafael Tolosana Calasanz (University of Zaragoza, SC23, ICPP23, ICPP24 Reproducibility Chair, and IEEE TPDS Associate Editor-in-Chief for reproducibility), Sascha Hunold (TU Wien, SC24 Reproducibility Chair), Brian Kocoloski (USC-ISI, co-PI for SPHERE), Tanu Malik ( University of Missouri, Columbia, SC21 Reproducibility Chair), and Christian Perez (INRIA, co-PI for SLICES) -- selected 10 presentations from 14 submitted proposals. The committee’s composition included a mix of practitioners positioned to define problems and platform providers positioned to contribute solutions. Presenters included Samuel Grayson (University of Illinois at Urbana-Champaign), Quentin Guilloteau (University of Basel), Triveni Gurram (Northern Illinois University), Kevin Kostage (Florida Gulf Coast University), Klaus Kraßnitzer (IST Austria), Ruben Laso (University of Vienna), Tanu Malik (University of Missouri, Columbia), Akhilesh Raj (Vanderbilt University), Adithya Raman (University of Buffalo), and Bogdan Stoica (The University of Chicago). Collectively, the committee members and presenters have chaired, authored, and/or reviewed artifacts at a diverse range of international conferences including Bench (2022, 2024), CCGrid (2025), Euro-Par (2024), EuroSys (2025, 2024), ICPP (2023, 2024), IEEE Cluster (2022), IEEE TPDS (2022), IPDPS (2025, 2022), SC (2024, 2023, 2021), USENIX OSDI/ATC (2023, 2024), USENIX FAST (2022, 2023, 2024, 2025), and PPoPP (2018). Their presentations \cite{repeto_presentations_hpc2024} spanned a wide range of topics -- from practical challenges in artifact packaging and evaluation, to specialized reproducibility concerns in areas like energy efficiency, visualization, and machine learning, while addressing both technical and procedural aspects of reproducibility in HPC. The program was anchored by a keynote address from Torsten Hoefler, known for his contributions to reproducibility in HPC research \cite{hoefler2023reproducing,hoefler2015scientific,ben2019modular}.

The technical program was divided into morning and afternoon sessions, each moderated by reproducibility chairs from SC (Sascha Hunold and Tanu Malik). Each presenter delivered a presentation followed by a presenter panel discussions, where presenters engaged with specific questions about critical reproducibility topics (see \textbf{\hyperlink{appendixc}{appendix C}}) with moderators facilitating audience participation. This discussion format allowed many participants to bring valuable perspectives from their experiences publishing research on reproducible practices in systems and HPC (e.g., Beth Plale \cite{plale2021reproducibility,plale2021transparency}, Tanu Malik \cite{plale2021reproducibility,kamath2024fair,ahmad2022reproducible,malik2022expanding}, Michael Heroux \cite{heroux_improving_cse_2011,osti_1481626,heroux2019trust}, and Florina Ciorba \cite{guilloteau2024longevity,guilloteau2024reproducibility}) or their roles as AE authors and reviewers at a diverse set of conferences. The workshop also benefited from the participation of two presenters who had recently completed the Summer of Reproducibility program in 2024 \cite{summer_of_reproducibility}. Regular breaks throughout the day provided opportunities for informal community discussion and networking.

The workshop's impact extended beyond its single-day format through a  BoF session held during the SC24 conference \cite{sc24_bof_practical_repro_hpc}. This follow-up session, led by several workshop organizers including Sascha Hunold, Daniel S. Katz, Kate Keahey, Jay Lofstead, and Tanu Malik, summarized the workshop's findings and used them as a springboard for broader community engagement with SC24 attendees. The BoF format allowed for interactive discussions on key themes identified during the workshop, effectively expanding participation and extending the practical reproducibility conversation to the wider SC24 audience.

Overall, the workshop attracted significant interest from the HPC community, with 56 registrants and approximately 40 active participants at any given time (amplified by an additional 20 attendees at the BoF). Through targeted outreach efforts via multiple channels -- including HPC conferences that conduct reproducibility initiatives (facilitated by Chameleon's widespread use as a supporting platform), the Better Scientific Software (BSSW) blog \cite{bssw_repro_workshop_2024}, InsideHPC events \cite{insidehpc_site}, HPC industry groups (i.e., HPC.social \cite{hpc_social_site}), and the Chameleon user community -- the workshop's website garnered over 2000 unique visitors in the months leading up to the event. The organizing committee made a concerted effort to engage thought leaders in reproducibility research, successfully bringing together representatives from complementary infrastructure projects such as the Chameleon \cite{chameleon_project}, SPHERE (Brian Kocoloski) \cite{sphere_testbed}, SLICES (Christian Perez) \cite{slices_ri}, and FABRIC testbeds (Paul Ruth) \cite{fabric_testbed}.

Subsequently, the organizing committee initiated a structured process to compile in this report the challenges and recommendations that the community surfaced during these events. First, the committee collected all workshop notes, presentation materials, and discussion outcomes from the workshop and BoF. The lead authors then prepared an initial draft, which was then discussed and significantly extended in discussion among the  authors. After incorporating insights from this discussion cycle, the committee circulated the revised draft to another 60 stakeholders from the broader HPC reproducibility community -- including all workshop attendees -- for public comment (March 17–28, 2025). Overall, the report received 101 comments from 15 contributors across the two discussion periods. We would like to particularly acknowledge contributions of Paul Brunk (The University of Georgia), Quentin Guilloteau, Triveni Gurram, Michael Heroux (ParaTools, Inc., St. John's University), Daniel S. Katz (National Center for Supercomputing Applications, University of Illinois Urbana-Champaign), Brian Kocoloski (SPHERE co-PI, USC-ISI), David Koop (Northern Illinois University), Klaus Kraßnitzer, Ruben Laso, Ken Raffenetti (Argonne National Laboratory), Paul Ruth (RENCI, UNC-Chapel Hill, FABRIC PI), and Bogdan Stoica (The University of Chicago). The substantial feedback received during both comment periods significantly shaped the final content of this report, which reflects both the workshop and BoF discussions as well as insights from referenced literature and community expertise. This multi-stage review process ensured comprehensive coverage of the challenges and recommendations while maintaining alignment with community perspectives.

\section{Background: Definitions and Incentives}
\label{background}

\subsection{Terms and Definitions}

\textbf{Experiment}: A computational study conducted on computing resources that includes code, data, environment, workflows, procedures, and results designed to test hypotheses or demonstrate techniques. Experiments can usually be seen as composed of three stages as defined in \cite{keahey2023three}: (1) the creation of \emph{experimental environment} or topology: the allocation, configuration, and orchestration of resources in which the experiment will execute (e.g., ``create Linux cluster with a distributed storage system"), (2) \emph{experiment body}, i.e., the actual execution of experimental actions (e.g., benchmarking the created environment), and (3) \emph{data analysis and presentation}. Experiments may emphasize or de-emphasize some of those stages; for example, they may skip the creation of an experimental environment if, e.g., the objective is to discover new properties of existing environments (e.g., a specific datacenter configuration). 

\noindent
\textbf{Experiment Authors}: Researchers who create experiments and package them for reproducibility.

\noindent
\textbf{Experiment Reviewers}: Researchers who attempt to reproduce experiments. This reproduction might be done as part of an AE initiative (where a formal review takes place) or for personal researcher use (where no review is provided).

\noindent
\textbf{Artifact Description (AD)}: A comprehensive document that details all components necessary to reproduce an experiment, including software dependencies, hardware requirements, configuration parameters, expected outputs, and step-by-step execution instructions. The AD serves as a guide for anyone attempting to replicate the research results.

\noindent
\textbf{Artifact Evaluation (AE)}: A formal process, typically organized by conferences or journals, where independent reviewers attempt to reproduce experiments based on ADs and documentation provided by the authors. The evaluation assesses reproducibility, usability, and completeness of the research artifacts, often resulting in badges or certifications that recognize reproducible research.

\noindent
\textbf{Artifact}: The collection of research objects including code, datasets, environment, configuration files, and documentation that together enable the reproduction of experimental results described in a research paper.

\noindent
\textbf{Reproducibility}: The ability of an independent researcher to obtain consistent results---the same or equivalent results---by following the documented experimental procedure using the same methods, code, and data as the original experiment.

\subsection{Reproducibility Badges}

Badges are visual indicators awarded to digital artifacts of research papers that have successfully demonstrated specific aspects of reproducibility. Organizations, i.e., ACM \cite{acm_artifact_badging}, the National Information Standards Organization (NISO) \cite{niso2021rp31}, and OSF \cite{osf_badges}, have developed standardized badging systems to recognize and incentivize reproducibility efforts across publications. Many systems and HPC conferences use these systems, which serve as an important mechanism in the scientific community:

\begin{itemize}
    \item Recognition: They publicly acknowledge authors who invest extra effort in making their research reproducible.
    \item Incentivization: They encourage good practices in research documentation and artifact preparation.
    \item Signaling: They help readers quickly identify research with verified reproducibility claims.
\end{itemize}

\noindent
Common badge types in the ACM system and similar badging initiatives include:
\begin{itemize}
    \item Artifacts Available: Indicates that the artifacts associated with the research are publicly accessible in a permanent archive.
    \item Artifacts Evaluated -- Functional: Certifies that the artifacts can be executed and perform as described.
    \item Artifacts Evaluated -- Reusable: Signifies that the artifacts are well-documented, easy to reuse, and can be adapted for other research.
    \item Results Reproduced: Confirms that an independent team has successfully reproduced the central results using the artifacts.
    \item Results Replicated: Indicates that an independent team has verified the results by developing their own artifacts based on the paper's description.
\end{itemize}

These badges typically appear on the first page of published papers and in digital libraries. The existence of these badging systems provides context for the challenges and recommendations discussed in the following sections, as they represent the formalized standards against which reproducibility efforts in HPC are currently measured.
\section{The Challenges}
\label{challenges}

The workshop discussions revealed and characterized many challenges in practical reproducibility ranging from packaging and review to support for the review process. This section summarizes that categorization in as much detail as we could capture to provide a background and justification for workshop recommendations. 

\subsection{Challenges of Experiment Packaging and Review}
\label{experiment-package-challenges}

The objective of this section is to characterize and explain the technical challenges in packaging computer science experiments in systems and HPC. We seek to define \textbf{what makes experiment packaging and its review challenging and time-consuming}, the most significant obstacles to practical reproducibility. We break these challenges down to address various aspects of complexity and seek to represent both the author’s and the reviewer’s perspective.

\textbf{Completeness of artifact description}: Artifacts are challenging to package largely because authors find it hard to articulate all the details of an experimental configuration with the required level of accuracy. This problem is particularly acute when the authors are working with a private/closed environment that they themselves have not explicitly configured \cite{keahey2020silver}; in this case they may not even be aware of some of the critical details. Another reason is the sheer volume of detail that often needs to be communicated exactly to ensure smooth reproduction: unlike in a paper, there is a very high likelihood that an innocuous typo or underspecification will make the artifact hard or impossible to reproduce without significant debugging (see also, e.g., \cite{krassnitzer2025autoappendix}). Thus, there is a higher standard for artifact accuracy and “getting things exactly right” than in a paper. Lastly, much information is simply lost in communication: often artifact descriptions are incomplete or make assumptions about “common knowledge” -- that is not in fact common. 

Some common issues that make artifact descriptions incomplete (or impact their completeness) are as follows:

\begin{itemize}
    \item Lack of description of the needed hardware and/or implicit or underspecified assumptions about the hardware.
    \item Missing software and data dependencies (unspecified, wrong/incompatible versions, bugs introduced through the package management software and others.
    \item Failure to note when proprietary code/hardware/licenses are used or how to obtain them -- or proxies/emulators that can replace that functionality with accessible alternatives.
    \item Failure to note restrictions on component accessibility (e.g., specified dataset may not be accessible to reviewer, too large to store conveniently, or involving too lengthy a download).
    \item Incomplete description of experimental environment, e.g., lack of environment variable specifications required (performance tuning settings often read from an environment variables).
    \item Poor version control of artifact sources (broken links; commits not fixed/specified; incomplete repositories due to modifications in local copies).
    \item Code quality/bugs that often manifest themselves only in a slightly different environment.
    \item Failing to include control strategies for obvious errors.
    \item Lack of management of expectations about experiment longevity (a hard problem in general \cite{guilloteau2024longevity}) -- unmaintained software decays and a reproducible experiment may simply get out of its “best before” date -- a fact that may not be clear until after much effort has been invested in a reproduction.
\end{itemize}

\noindent
These challenges also contribute to the debate about the role of AE reviewers: are they meant to ``debug" the artifact or simply evaluate the artifact? How much effort should the reviewer put in to reproduce/debug an artifact? Challenges of artifact quality was also discussed in \cite{malik2024fair}.

\textbf{Finding the right specialized/custom hardware}. HPC/systems experiments often need unique/bleeding-edge hardware and often at large scales. Finding such resources is hard for the purpose of the original experimentation -- let alone to reproduce experiments, as reviewers may not have the same level of access as the author. This problem is compounded by that fact that, if the same resources are not available, the reviewer has to \emph{approximate the resource requirements in terms of different resources} which creates challenges of its own. 

The need for \emph{unique hardware} is specific to computer science experimentation where particular features are either an \emph{object of experimentation} or \emph{needed to obtain data about the object of the experiment}. Examples of the former include experimentation with a specific type of architecture or custom network that may require matching an author's hardware specifications at not only the vendor level, but at the level of microarchitecture, model number, perhaps even firmware and OS device driver version, especially for GPUs and other accelerators. An example of the latter is power monitoring infrastructure: energy efficiency studies (i.e., \cite{raman2024energy}) often rely on dedicated power measurement tools (at node-level granularity) that are unavailable in standard HPC environments. Lastly, \emph{non-disclosure requirements} for testing hardware can also be a problem in which case they typically require vendor intervention.

Another dimension of complexity is the need for \emph{large scale} that could represent a prohibitive amount of resources. Examples of the need for large scale include experiments on node- and memory-intensive workloads that may require potentially \emph{hundreds or thousands of nodes}; experiments that store large intermediate datasets or ones that probe the limits of extreme scale storage and I/O protocols that typically have \emph{high storage and I/O demands}; or experiments requiring long runtimes that require \emph{longtime availability} of the resources they use. Proxies like SimGrid \cite{simgridproject,CASANOVA2025103125} and NetLab \cite{netlab2023} exist to mitigate large-scale resource needs, but remain underutilized due to two factors: (1) researchers with abundant access to large-scale resources often lack awareness of these alternatives, and (2) the significant time investment required to learn and properly configure these simulation environments creates an additional burden on researchers already facing publication deadlines.

Using \emph{open research infrastructure or shared infrastructure} has the potential to solve the problem, but currently does not. This is partly because \emph{not all shared platforms support unique features} such as access to power monitoring for energy efficiency experiments -- or are configured such that those features are available to local users but not to everybody. To some extent, it is also because \emph{shared infrastructure is currently not widely used by authors} \cite{krassnitzer2025autoappendix,guilloteau2024longevity}; in the best case, this means that the reviewer has to invest in porting the work to a shared platform which adds extra effort -- in the worst case, the experiment may contain hidden dependencies on local hardware or its configuration \cite{keahey2020silver}. Lastly, access to hardware that is much in demand may require an advance reservation and some wait time (see, e.g., \cite{malik2024fair}); this is \emph{not currently well supported by the short timelines of the artifact review processes}.

\textbf{Quality of artifact description}: Even artifacts that are in principle complete in their description may be needlessly difficult to reproduce due to clumsy packaging, poor research practices, or insufficient communication to the reviewer. These missteps are easy to overlook for the author who is intimately familiar with the artifact -- but have a disproportionate impact on the reviewer, who is not. While in principle they do not prohibit the reviewer from reproducing the experiment, they make it much harder to reproduce, increasing the reviewer’s cost so that while the experiment may be possible to reproduce, it will not be possible to reproduce \emph{practically}. The discussions at the workshop specifically called out the following bad practices:

\begin{itemize}
  \item Experiments that are not organized into well-defined components or phases such as establishment of the experimental environment, execution of the experiment body, data collection, and data analysis -- where possible, with clear checks on whether each phase was completed successfully.
  \item Failing to provide runtime messages and software-defined checkpoints.
  \item Hardwiring changeable qualities (variables/parameters). An example is returning an error when a random hostname is tried, as opposed to asking reviewer to define changeable qualities/ constants like hostnames up front, perhaps via command line argument or environment variable.
  \item Not managing expectations as to the time commitment to reproducing an experiment (or its components). Having a realistic expectation helps the reviewer plan -- and also may and provide an indication of whether/if the reproducibility attempt is going well.
  \item Not including teardown commands (may have an impact especially if the experiment is tried multiple times).
  \item Not clearly articulating the “reproducibility condition” -- or providing at least some discussion on when the experiment may be considered reproduced. While this is a hard problem in general (also see below), clarity on this point not only aids reproducibility, it also helps define the result.
\end{itemize}

All in all, these challenges point to the need for a style guide -- just like writing papers, i.e., communicating research results in a clear and persuasive way, needs to be learned, so does packaging experiments such that they are reproducible. 

\textbf{Establishing experimental environment}: Experimental environments range from relatively simple, e.g., a single compute node with specified configuration, to more complex, e.g., ``MPI cluster" specifying configuration for multiple compute nodes (which involves the need to get exact versions of libraries and tools), to very complex distributed system topologies requiring configuration of compute nodes, networks, and storage elements. Creating an experimental environment is a complex task: it typically involves the installation of multiple layers of firmware and software, with the right versions and configurations, securely exchanging and integrating information generated at runtime (e.g., dynamically assigned hostnames and security tokens), and setting the right configurations. Because of this complexity, establishing a correct environmental environment tends to be the most complex and error-prone part of reproducing an experiment so that it is usually critical to automate it. For many experiments, it is also the most time-consuming one. 

The common complexity elements and trade-offs discussed at the workshop included:

\begin{itemize}
  \item Automating the complex environment build process and its contextualization \cite{keahey2008contextualization}: While there are many tools/methods available (Nix(OS) \cite{dolstra_nix}, Guix \cite{guix}, Spack \cite{Gamblin_The_Spack_Package_2015}, Parsl \cite{babuji19parsl}, Heat \cite{openstack-heat-2024}, Terraform \cite{terraform}, etc.), there are no clear winners and it is non-trivial to pick the most suitable tool for the problem -- or the platform -- and the author’s choice may not be the reviewer’s. The problem is compounded by the fact that many of those tools are geared towards operators rather than end-users so that using them will require a high level of skill.

  \item Snapshotting versus ``recipe": For reproducibility in the short term, environment packaging is often best done via snapshotting images and containers configured with the experimental environment -- but such snapshotted images effectively replace a “recipe” on how to reproduce them (they are a shortcut) and are hard to maintain or reproduce in the long term. Providing and executing a recipe has more potential for longevity but may take a long time for the author to prepare and is generally more error prone.

  \item Declarative versus imperative styles trade-off \cite{anderson2019case}: declarative styles of packaging (e.g., orchestration) ask a user to specify the desired state of the system and execute that state. However, they are transactional; if they fail the user has little insight into the process. In addition, the same state can be achieved in multiple ways, each associated with different side-effects, which may interfere with the objective to reproduce. Imperative styles, on the other hand, are non-transactional but less user-friendly as they focus on describing how an environment is created rather than the desired state. Finding the right balance between these approaches is a significant challenge.

  \item Time: even if the process of creating such environments is clear and automated, the builds can take a long time -- if they need to be repeated (as they often do to address some hidden incompatibilities) these long times have a cumulative effect on the time of the review, significantly contributing to the cost of reproducibility.
\end{itemize}

\textbf{Finding the “reproducibility condition”}: It is often hard to unambiguously decide whether an experiment has been reproduced or not, as it requires distinguishing between meaningful and trivial or incidental changes (e.g., whether a graph shows the same ``trend" can be very much in the eye of the beholder). Inherently part of the problem is that the current practice may sometimes expresses the results indirectly in terms of, e.g., a time series rather than properties of the time series (though presumably the properties can be seen on the time series). Reproducing performance results presents unique difficulties as performance metrics are highly sensitive to hardware specifications, system configurations, and environmental factors \cite{hoefler2024performance, plale2021reproducibility,sedghpour2024artifactevaluation}. Small variations in these conditions can lead to significant differences in results, making it difficult to establish clear acceptance criteria for successful reproduction. Experiments often fail to include statistical methods like error bars, confidence intervals, or other quantitative measures that would help define when a result has been successfully reproduced within acceptable margins of error, especially when comparing trends rather than exact values \cite{hoefler2015scientific}. Difficulties and strategies for deciding whether an experiment is in fact reproduced or not reproduced by direct comparison were heightened in both the keynote \cite{hoefler2024performance} and one of the presentations \cite{koop2024visualization}.

\textbf{Cost}: Reproducibility may incur a significant cost both in terms of human effort (of both authors and reviewers) and experimental resources (hardware or consumed energy). In both cases, this cost is multiplied on the review side if the artifact is not well packaged in that the same potentially costly actions will have to be repeated many times before achieving a successful reproduction expending both human and computer resources. In both cases also, in addition to direct cost there is also an opportunity cost: the reproducibility work will likely be done at the expense of new research. 
 
While this cost varies greatly depending on the complexity of the experiment, several helpful studies sought to provide a rough estimate. A comparison brought up at the workshop estimated that reviewing one artifact can take more time than reviewing 6 papers. In a survey administered to participants after the SC21 AE, the chairs found that reviewers spent an average of 13 hours per artifact assessing the description and reproducing the experiment (weighted by number of artifacts each reviewer evaluated); reviewers were assigned on average 3.5 artifacts to review, totaling about 40-50 hours of commitment per reviewer \cite{malik2022expanding} (see also discussion in \cite{collberg2016repeatability}). A recent longitudinal study of 744 ML-security papers found reviewers needed \textbf{approximately 10,000 CPU-hours and 8 person-years} to complete direct-reproduction attempts of all included artifacts, even after code was available \cite{olszewski2023get}.

\textbf{Longevity}: After much care and effort has been expended on creating and validating artifacts, those artifacts may be of limited usefulness due to their limited lifetime. Artifacts may have a strong dependency on unique hardware features, which may no longer be available after a certain time, and have a natural time horizon that would require special arrangements to extend \cite{guilloteau2024longevity}. Even when this is not the case, most software decays without maintenance and this decay will increase the level of difficulty of reproducing artifacts. Specific challenges of "workflow decay" have been identified in research, where scientific workflows become unusable over time due to changes in the underlying services, dependencies, or lack of sufficient metadata to understand their purpose \cite{hettne2012best}. In some cases, this decay may lead to security vulnerabilities being introduced into the artifact; this means that the artifact, although still reproducible, in practice is not reproducible in the same way (i.e., might require special conditions for reproduction). In general, different longevity is suitable for different artifacts: results that are very specific to the technical ecosystem of the moment, while potentially impactful in their time, may be of limited general value.

\textbf{Findability}: Many conferences and journals have adopted reproducibility practices, dedicating substantial effort to the process; as a result, their accepted papers can earn reproducibility badges. Those papers are indexed in digital libraries such as ACM Digital Library (DL) \cite{acm_digital_library} and IEEE Xplore \cite{ieee_xplore}, but the reproducibility artifacts are not (ACM DL provides limited support for filtering papers that have been awarded reproducibility badges). Such collections of artifacts as exist are not searchable \cite{sysartifacts2024}. This makes it almost impossible for artifacts to be useful after the AE initiative is over, because they are not easily findable -- a problem compounded by the fact that papers are stored in the digital library platform, whereas most artifacts are hosted on platforms like GitHub. Even if artifacts were indexed, just locating them would be of limited value if they are not associated with resources on which they can be run; conventional libraries address this problem for similar artifacts by providing the means of interpreting them (e.g., a library that stores microfilms might provide a microfilm reader). This provides an additional argument for packaging artifacts on open, shared resources -- preferably in a way that facilitates their execution.

\subsection{Challenges of the Artifact Evaluation Process}

This section seeks to answer the question of \textbf{what makes organizing AE initiatives challenging}. We break it down into various aspects below. We seek to represent both the author’s and the reviewer’s perspective.

\textbf{Incentives and recruitment}. The cost of reproducibility, detailed across the discussion of different challenges in the previous section, raises the question of incentives: who benefits from reproducible research, in what ways, and to what extent -- so that the investment of time in either packaging or reproducing experiments is balanced by benefits that make it worthwhile? Simply put, establishing a viable cost/incentive balance is critical to sustaining long-term interest of reproducibility as a scientific practice.

For the author, the cost of reproducibility consists not only of just packaging an artifact -- but ideally, packaging it such that the reviewer needs to invest the least possible time in its reproduction and such that it can be sustained and have impact beyond the AE initiative (see Section \ref{experiment-package-challenges}). Artifact authors are incentivized by conference requirements to submit reproducible artifacts as well as public recognition which comes in the form of badges, awards, and similar vehicles. These vehicles present their own challenges as they currently present a ``flat" acknowledgment, i.e., do not reflect factors such as the quality, level of effort required, or innovation that a packaging represents. Ultimately, the authors should be incentivized by impact -- if reproducing their research leads to new research that is directly derived from reproducing their results and acknowledges them. However this “virtuous cycle” is hard to get off the ground, for two reasons. First, it is hard to talk about impact in a meaningful way, given the fact that currently artifacts are rarely reproduced outside of the AE initiatives (thus, practical reproducibility is critical to solving the incentive problem sustainably). Second, there is a lack of a ``recognition ecosystem," consisting of accepted methods of acknowledging the role of specific reproducible artifacts in initiating or advancing new research; translating such acknowledgments into credit to authors (similarly to what, e.g., citation counts for papers accomplish today); and using this credit to advance the author’s career and scientific standing. Less directly, reproducibility can pay off for authors indirectly through fostering good research practices, amplifying their impact \cite{markowetz2015five} -- though arguably in many cases this indirect benefit is valued less than the direct benefit of investing in new research instead.

If incentives for authors are problematic, incentives for reviewers are worse. \textbf{This means that reviewers for reproducibility initiatives are exceptionally difficult to recruit}. Reproducing experiments is challenging and takes a lot of time -- significantly more time than reviewing papers (see \textbf{Section \ref{experiment-package-challenges}}) -- and the AE structure and approach that a conference adopts can influence the amount of work reviewers do to evaluate them. For example, while open, collaborative processes with authors has many benefits, it is also likely to extend the amount of time involved. In addition, the fact that each AE process can be different for reviewers requires them to re-learn the process every time, introducing more friction, and reducing the attractiveness of participating across conferences (this can be true within the same conference series across years: e.g., SC Chairs change their processes each year). At the same time, while authors at least stand to receive badges for reproducible artifacts reviewers don’t receive any credit or recognition for their involvement and for the work they do \cite{hunold2024evolving}. As with the author, ultimately reviewers may be incentivized by the inherent learning opportunity that reproducing artifacts represents; however given the still poor quality of artifacts and lack of a viable “recognition ecosystem” at present the drawbacks often outweigh the benefits.

\textbf{AE timelines don’t give authors enough time or guidance to package experiments}. In a typical AE initiative timeline, authors may be given as little as 2 weeks between notification of acceptance and the deadline for submission of their artifacts \cite{Guilloteau2024}. For example, SC24 notified authors of paper acceptance on June 14, 2024 while the artifact submission deadline was on June 28, 2024 (14 days); USENIX OSDI/ATC24 notified authors of paper acceptance on April 30, 2024 while the artifact submission deadline was on May 12 (12 days). While in principle authors could package artifacts prior to acceptance, this is usually not the case for the following reasons: authors may be reluctant to release complete artifacts or code before paper acceptance decisions due to intellectual property concerns; not wanting to invest effort in the face of publication uncertainty (especially given that reviews might cause the need to reposition the results); or might not receive sufficient information on packaging requirements from the AE initiative before paper acceptance -- especially challenging for authors who are packaging artifacts for the first time and need such help.

\textbf{AE timelines don’t give reviewers enough time/guidance to review}. In a typical AE initiative timeline, reviewers may be given as little as 2 weeks to review artifacts. For example, during SC24 reviewers were assigned and granted access to final artifacts on August 16, 2024, and badge decisions are issued on August 23, 2024 (13 day window); during USENIX OSDI/ATC24 reviewers were able to access artifacts on May 15, 2024 with decisions due by May 31, 2024 (16-day window). These timelines mean that although artifacts require significantly more time and effort to review on average than papers do (see \textbf{Section \ref{experiment-package-challenges}}), AE reviewers are typically given \textbf{less} time to evaluate artifacts than paper reviewers have to evaluate papers. The problem is compounded by the fact that, even if authors do use open infrastructure, the reviewer may have to take time to learn how to use that infrastructure; the resources may not be available on-demand (major testbeds like Chameleon and CloudLab arbitrate access to most popular resources via advance reservations but it may take up to a few weeks to gain access to resources); and the artifacts are currently often not of sufficient quality to make the review process seamless. Some venues attempt to mitigate time constraints by assigning multiple reviewers per submission (e.g., ICPP assigns 3 reviewers per artifact), but this approach demands more computational resources overall and creates coordination challenges.

\textbf{The AE process is in the process of being invented}. The AE review model is in the process of being invented which results in much confusion among authors and reviewers participating in multiple AE initiatives. Further, most of the current AE initiatives follow the model set by paper reviews but there are signs that it is not entirely appropriate. There is tension between different models of AE: closed versus open, limited feedback versus collaborative improvement, strict deadlines versus extended timelines. The workshop discussion centered on the following questions:

\begin{itemize}
    \item What is the objective of the AE: should it be to merely evaluate an artifact or to improve it? If improve, how do we avoid extreme situations where the reviewers end up doing a significant amount or even most of the packaging work in practice?
    
    \item What is the role of a reviewer: should it be to debug the artifact, improve it, or just evaluate it? How do we set the boundaries between those goals? The reviewers may end up significantly contributing to the final form of the artifact; should we credit their contribution? (possible only in open reviews)
    
    \item If an artifact requires a lot of support from the author to reproduce should it be considered reproducible in practice? How can the review process best ensure that the review in fact results in an improved artifact, rather than a one-time successful reproduction?
\end{itemize}

Different answers to those questions from different conferences result in different approaches which has different implications for the quality of artifacts, reviewer workload, and the educational value of the process.
\section{Recommendations}
\label{recommendations}

\subsection{Recommendations for Packaging Experiments}

\hypertarget{rec1}{
\textbf{Recommendation 1 (authors, reviewers)}: Both authors and reviewers should \textbf{use (and when possible help improve) the artifact packaging checklist in \hyperlink{appendixa}{appendix A}} to ensure the completeness of artifact description. Not every item in the checklist may be applicable to every experiment; rather the checklist should serve as a guide to be used at the author’s discretion when preparing an artifact for sharing. While experiment authors can use the checklist for packaging artifacts, the reviewers can use it to evaluate a package and form a judgment on the quality of its preparation for reproduction. Since learning about what may be required is an ongoing process, \textbf{Section \ref{contributions}}} also contains instructions on how the checklist could be improved.

\hypertarget{rec2}{
\noindent
\textbf{Recommendation 2 (authors, reviewers)}: Both authors and reviewers should \textbf{use (and when possible help improve) the experiment style checklist in \hyperlink{appendixb}{appendix B}} to ensure good practices of artifact description. Even when the artifact description is complete, how experiments are structured and documented significantly impacts their reproducibility and in particular its associated cost -- and is particularly critical to practical reproducibility, i.e., ensuring that those experiments can be used repeatedly for scientific exploration rather than for just a one-time reproduction. Where ensuring AD completeness is a necessary condition for artifact reproducibility, following good practices represents a sufficient condition that will help the experiment be useful beyond an AE initiative. These style guidelines summarized in \textbf{\hyperlink{appendixb}{appendix B}} -- such as clearly articulated experiment structure, providing intermediate consistency checks, describing the expected time to reproduce, or divulging clear reproduction criteria -- help ensure good reviewer experience while improving artifact quality and reproducibility.
}

\hypertarget{rec3}{
\noindent
\textbf{Recommendation 3 (authors)}: Authors could try a ``sample evaluation" of their artifact from the AD. If applicable, this should ideally happen on the target hardware (or generally other hardware than the one the experiments were originally conducted on). In a study of SC24 artifacts on Chameleon \cite{krassnitzer2025autoappendix}, one of our presenters found that most authors had never run their artifacts themselves and thus did not realize what was missing from the reviewer's perspective.
}

\hypertarget{rec4}{
\noindent
\textbf{Recommendation 4 (authors, reviewers, organizations)}: Where possible \textbf{use open shared infrastructure}, to package experiments. In particular, this includes testbeds specifically adapted for computer science experimentation such as NSF-supported testbeds including Chameleon \cite{chameleon_project}, CloudLab \cite{Duplyakin+:ATC19}, SPHERE \cite{sphere_testbed}, FABRIC \cite{fabric_testbed}, and the PAWR testbeds \cite{pawr2025} -- as well as international testbeds such as SLICES \cite{slices_ri}, EGI \cite{egitestbeds}, and EOSC SU Node \cite{eosceunode}. Deeply reconfigurable infrastructure like Chameleon and CloudLab cover the greatest set of experiments by providing access to unique hardware and bare metal reconfigurability, which means the environment is fully under the experimenter’s control --  though they may not always satisfy the requirement for extreme scale. At the most basic level, using this type of testbed means that \textbf{reviewers gain access to a diversity of deeply reconfigurable resources} so that at least that obstacle to review is eliminated. Further, if authors package artifacts for shared infrastructure, this both \textbf{increases the likelihood that a reviewer can reproduce the experiment} \cite{keahey2020silver} and \textbf{lowers the time demand on the reviewer}. Digital libraries like Trovi \cite{trovi2025} can further facilitate it by providing a seamless integration between experiment artifacts and hardware. To facilitate such usage, AE organizers are encouraged to \textbf{work directly with testbed providers} to ensure availability of information and training and coordinate resource availability at times of intense demand as needed.
}

\noindent
\textbf{Recommendation 5 (authors, community)}: As needed, package emulators or other tools (e.g., SimGrid \cite{simgridproject} and NetLab \cite{netlab2023} ) providing acceptable substitutes for representing infrastructure such as sufficient scale. Good quality emulators with widespread appeal are also something that may require further investment; they may need to be developed by the community to represent specific problems.

\noindent
\textbf{Recommendation 6 (resource providers, community, organizations)}: Consider initiatives \textbf{making unique resources available for reproducibility initiatives on a temporary basis} at specific times throughout the year. This could take the form of a “reproducibility month,” or a virtual ``open house," for specific resources when a unique resource or its portion is set aside for reproducibility access for a limited time based on prior reservations. Making some part of the system available for reproducibility could also be a competitive element of infrastructure proposals. Resources could be exposed in the way they are typically operated (e.g., by submission to a batch scheduler), though an interesting option would be to make them temporarily available through an interface supporting deep reconfigurability, as this would support a broader set of experiments than these resources normally support. The capability to do this, as well as to broker access to resources configured in this way, to some extent exists through systems like CHI-in-a-Box \cite{keahey2022chi}, though improving its efficiency for ephemerally managed resources would require further refinement. Alternatively, resource providers should also consider making specific allocation types for researchers making use of reproducibility (such as, e.g., \cite{chameleon_daypass}) in conjunction or instead of “open house” initiatives. AE initiatives and resource providers could also \textbf{work with vendors to ensure that NDA requirements for testing hardware are managed} for authors/reviewers.

\noindent
\textbf{Recommendation 7 (authors, community)}: Not all scale problems can be addressed through resource availability or use of emulators and simulators; where this is the case, authors and reviewers may try to \textbf{scale down the experiment instead}, or employ other problem-based techniques. Sound methods for scaling down an experiment, might include, e.g., using Cgroups \cite{verma2015large,linuxcgroupsv2}, to throttle hardware capabilities can recreate conditions that would normally only exist at large scale. This approach solves multiple problems: the hardware requirements are likely to represent a much lower barrier; the cost of reproducibility (in both human and infrastructure terms) will be lower; and in some cases this may even better capture the result. On the other hand, applying this technique may increase packaging time for the author. Compiling a list of recommendations of how to scale down different types of experiments would be an important methodological contribution by the community that would greatly facilitate the use of this technique -- especially if it could become part of experimental methodology training. Another idea -- that does not reproduce the problem \emph{per se} but aids in verification of results --  are meta-computations \cite{heroux2019trust}, i.e., the idea that by performing some computations that probe for correct behavior based on an understanding of the problem being solved and the algorithms being used, we can better ensure a correct answer. Meta-computations can also be useful in verifying the stages of, e.g., establishment of an experimental environment. An intriguing idea is to explore generating them based on AI techniques.

\noindent
\textbf{Recommendation 8 (community)}: Invest in developing \textbf{end-user focused tools and methodologies for repeatable environment packaging and re-establishment} as well as \textbf{teaching/training on how to package environments} \cite{fund2023need}. It is essential that such tools recognize ease of use (of both packaging and re-enactment) as a key cost lowering concern. It is also important that open research infrastructures support the necessary underpinnings for such automated environment building by exposing programmatic interfaces that allow users to capture the process of environment building in a programmatic and thus repeatable way.

One of the discussed ideas to resolve the ``snapshotting versus recipe" trade-off (see \textbf{Section \ref{experiment-package-challenges}}) was to develop and standardize approaches for \textbf{capturing both the environment and its build process} -- containers should include detailed, version-controlled build manifests, and when using package managers, specific version pinning should be enforced rather than defaulting to latest versions. This is currently not always easy to do, but extensions or additions to the existing tool ecosystem would bring it within reach. This may include developing tools to automatically generate build manifests from existing environments, capturing installed libraries and drivers with their versions.

To address the “declarative versus imperative” packaging styles trade-off, one of the recommended strategies is to use a \textbf{hybrid approach that uses declarative approaches for stable, well-understood components and imperative scripts with extensive error checking and logging for complex or custom components}. However, not all experiments fall neatly into a combination of these categories; regardless of approach, the overriding consideration is to ensure that the process is transparent and provides a clear indication of success or failure at each stage so that the experiment can be built incrementally.

Another key idea was to \textbf{create libraries of “experiment patterns”} \cite{Keahey2022}, pre-packaged versions of popular experimental environments representing e.g., virtual clusters with popular tools (MPI, SLURM) or installation of popular packages that will facilitate experiments (see discussion in \cite{krassnitzer2025autoappendix}). A pre-packaged environment of this kind can form the basis of a whole family of experiments relating to a specific configuration. To some extent, the movement towards such experiment patterns is already present; for example, the existing testbeds (e.g., CloudLab and Chameleon) maintain image libraries or experiment notebooks capturing popular environments. Building on this idea, providers of tool packages that are suitable for experimentation (e.g., MPI implementations) could support a packaging of their tools in the environment format suitable for deployment on relevant testbeds or commercial clouds; this would aid in dissemination of the tool on one hand and simplify the task of experiment packaging and review on the other.

Lastly, there is a high potential for recent advances in AI to provide solutions in this space through combining natural language processing capabilities with programmatic interfaces to infrastructure: so long as the experimenter can provide an accurate description of an experimental environment of the kind that are currently often included in papers today. Approaches based on Large Language Models (LLMs) could be used to generate code and environment scripts for its deployment, effectively supporting reproduction of an artifact from an artifact description. At the same time, the attendees noted that AI-based code can add more complexity, so there are trade-offs to consider. It was also noted that the feasibility of these approaches will strongly depend on data and experience derived from working with large communities of users spanning many different experiment types.

\noindent
\textbf{Recommendation 9 (community)}: Invest in \textbf{developing tools, methodologies, and educational content aiding in the alignment analysis of reproduced results}. Three potential approaches were discussed at the workshop. The simplest one was direct comparison, i.e., by comparing resulting values or images and deciding whether the ``delta" is in an acceptable interval; this was illustrated in both \cite{hoefler2024performance} and \cite{koop2024visualization} together with the risks this approach engenders). Another option is to express results in ratios or aspects of what to compare and establish acceptable thresholds for differences by, e.g., using standard practices for reporting statistical uncertainty such as error bars and confidence intervals; this not only articulates the result more precisely but also defines what constitutes successful reproduction, especially for experiments focusing on trends rather than exact values. For performance-focused experiments especially, it is important to provide statistical methods and tools for validating results across different hardware environments, including techniques for normalizing results, establishing confidence intervals, and quantifying the impact of environmental variations on reported performance metrics. Further, an idea similar to ``meta computations" discussed above was to provide tools, in addition to the result, to analyze potential alignment of reproduced results -- such automated comparison could be especially valuable if the results are complex to analyze. Lastly, a similar function could be provided by building LLMs or other AI systems to assess disparities between reproduced results and original experiment results; strategies for developing those systems should be informed by the methods discussed above and would require significant data ranging across a large diverse set of experiments and are therefore best approaches within established testbeds. 

\noindent
\textbf{Recommendation 10 (community)}: Develop \textbf{tools, strategies, and incentives that can minimize the human and resource cost of reproducibility requirements}. With the current state of the field, the most effective strategy for reducing cost would simply be to improve the experiment packaging tools and practices (as well as their penetration within the community by, e.g., improving teaching) so that less human and compute resources are expended on experiment reproduction. In this context, we note that an experiment is usually packaged once (though there is additional maintenance cost incurred to ensure its longevity) -- but intended to be reproduced many times. The balance of effort and amortization of that effort between authors and reviewers will depend on effective community practices: if an experiment is packaged to be practically reproducible the chances that it will be reproduced many times increase; if it becomes a community practice to use experiments as a mainstream method of scientific exploration, the authors will be more likely to package their experiments better. Some of the experiment cost may be managed by strategies discussed above -- in particular, creating a minimal viable example/experiment that represents the result (this may increase the cost for the author but decrease both reviewer and resource cost) and using AI-based methods to streamline the parts of experimentation that are currently particularly challenging.

\hypertarget{rec11}{
\noindent
\textbf{Recommendation 11 (authors, community)}: \textbf{Develop and use methods, best practices, and incentives to make artifacts not only reproducible but also maintainable}. The longevity of artifacts can be used by adopting best practices in packaging, e.g., through working to reduce the number of dependencies of an artifact (including hardware dependencies where possible) or packaging artifacts in a way that makes them not only reproducible but also easy to maintain. This is likely to introduce an extra cost on the part of the author, however it was also noted that it is often consistent with good methodology in terms of representing a generalized result. Security vulnerabilities, that may not always be easily eliminated from the experimentation process (e.g., if a dependency of a leading benchmark in a field acquires vulnerabilities) can usually be managed by deployment behind a bastion host \cite{Cooper2023} adding a minimal cost on the part of the reviewer. In general, the workshop participants agreed that ensuring artifact longevity is important but requires potentially significant extra work and should be recognized by additional form of credit in their form of longevity badges \cite{Guilloteau2024}.
}

\noindent
\textbf{Recommendation 12 (community)}: \textbf{We need digital libraries that will index reproducibility artifacts, stored in artifact-appropriate ways and integrated with platforms that support their execution}. Such libraries should allow users to search specifically for reproducibility artifacts related to a given topic or experimental methodology and, once a relevant artifact is found, the system should provide an intuitive way to link/navigate to the associated papers \cite{rosendo2023kheops}. Platforms like Google Colab \cite{GoogleColab}, Code Ocean \cite{cheifet2021promoting}, or Kaggle \cite{KaggleDatasets2023,KaggleCommunity2023} provide a promising concept in this space but integrate only with (a limited set of) commercial clouds and are frequently built on proprietary solutions, suitable only for only a narrow range of experiments. A viable digital library would index artifacts available in existing repositories (e.g., GitHub) and provide a recipe for integrating any infrastructure while representing consistent execution mechanisms for users. 

\subsection{Recommendations for Artifact Evaluation Initiatives}

\textbf{Recommendation 13 (organizations)}: \textbf{Sponsor -- or leverage existing sponsorships -- to incentivize a reviewer pool for AE}. For example, many conferences already invest in various initiatives that sponsor students through student volunteer initiatives, lower registration prices, or other forms of sponsorship. Some of those sponsorships could be made contingent on participation in AE reviews which would help incentivize the reviewer pool. Working with students in this capacity could be particularly rewarding as they have the most to gain from learning the experimental methodology lessons inherent in reviewing artifacts; benefit from reviewer training associated with initiatives (such as, e.g., access to testbeds, see \cite{fund2023need}); are generally most receptive to the adoption of new methodology; and can most benefit from adopting reproducibility practices in their careers. Another strategy, used by AI conferences such as ICML \cite{ICML2024guidelines}, is to organize an exchange whereby authors submitting artifacts are asked to review other artifacts; this not only helps recruit reviewers but also fosters appreciation for the ``other side" of the artifact review process. Some conferences established explicit forms of acknowledgment in the form of, e.g., reviewer awards (EuroSys recognizes committee members with Distinguished Artifact Evaluator awards \cite{eurosys2022artifacts}). AE organizers should work with conferences to align those incentives as much as possible.

\noindent
\textbf{Recommendation 14 (organizations)}: \textbf{Consider refining the available evaluation mechanisms and incentives}. The current incentives represent a flat structure and carry limited information about the quality/properties of the artifact. One suggestion to ameliorate the situation (and also potentially acknowledge reviewer’s contributions) was to attach AE reports written by the reviewers to the AD as they can provide much more information about the reproduction attempts (such AE reports were introduced by the AE SC24 initiative \cite{hunold2024evolving}). Another would be to introduce more nuanced acknowledgment mechanisms by refining the badging system and introducing new badges (see also \textbf{\hyperlink{rec11}{Recommendation 11}}).

\noindent
\textbf{Recommendation 15 (organizations)}: \textbf{Consider the use of AE checklists in the review}. In particular, the packaging and style checklists (\textbf{\hyperlink{appendixa}{appendix A}} and \textbf{\hyperlink{appendixb}{B}}) can be used for  both authors and reviewers. Specifically, for the reviewer, these checklists can serve both as a basis for a quick verification pass to ascertain that  the artifact meets the necessary conditions to be reproducible -- and then again offer categories for improving and providing a more considered assessment of the artifact.

\noindent
\textbf{Recommendation 16 (organizations)}: \textbf{Consider working with providers of open computer science testbeds in the organization of AE}. The use of open infrastructure, and in particular computer science testbeds, can help overcome the problem of finding resources for reproducibility as already noted in \textbf{\hyperlink{rec4}{Recommendation 4}}. In addition, those infrastructures can also support the AE with training and additional services that support the organizational structure of the AE; for example, Chameleon has supported multiple reproducibility initiatives in this way. Organizers should reach out to infrastructures to discuss such services as needed.

\noindent
\textbf{Recommendation 17 (organizations)}: \textbf{Adjust the timelines of artifact review process and, where possible, decoupling it from paper evaluation process}. The current AE timelines are too short for both authors and reviewers considering the need to prepare artifacts that adhere to high-quality standards (as per \textbf{\hyperlink{rec1}{Recommendation 1}} and \textbf{\hyperlink{rec2}{Recommendation 2}}) and test them (see \textbf{\hyperlink{rec3}{Recommendation 3}}) -- or to get trained in the review process, acquire resources, and ultimately review artifacts, potentially while consulting with authors or over multiple rebuttal phases. The timelines should fit the review process selected by a conference (see below) as well as its submission timelines that tend to differ significantly between conferences. Given that for some conferences they may be very tight, one proposal was that the artifact review should take place in the year following paper publication, with the idea that artifacts would be published at next year’s conference, providing for a generous time allocation for authors and reviewers. It was pointed out however, that the quality of artifacts may decay over the period of a year as may their timeliness. Decoupling AE from the paper submission could allow AE initiatives to focus more on improving the artifacts (which is a more collaborative, but time-consuming, process) as opposed to getting an artifact to work once.

\noindent
\textbf{Recommendation 18 (organizations, communities)}: Organizations should consider AE models that balance quality, efficiency, and incentives; and communities should invest in tools that support them. The workshop discussed the following approaches and their trade-offs:

\begin{itemize}
    \item \textbf{Closed, anonymous review} follows closely the current paper review process where reviewers remain anonymous; have no direct interaction with authors during evaluation; and can rely on paper review tools to manage the review. The advantages of this approach are that it maintains reviewer independence, controls bias, limits time commitment, and evaluates the artifact as a third-party researcher would experience it. The disadvantages are that it may result in rejection of artifacts with minor issues and forgoes the opportunity to improve the resulting artifact.
    \item \textbf{Open, collaborative reviews} means that reviewers are not anonymous and engage in direct, iterative communication with authors. The process resembles that used by software journals like the Journal of Open Source Software (JOSS) \cite{smith2018joss}, that are themselves based on code review process, and could use tools similar to or derived from rOpenSci \cite{Ram2019Community} and pyOpenSci \cite{pyOpenSci2023} to ensure that the artifact actually improves as part of the review. The advantages are that it has the potential to make the reviewer’s job easier (at least if the authors engage), has the potential to improve the artifact as part of the evaluation process resulting in higher artifact success rates and better quality, it has educational value for authors and reviewers, and supports reviewer recognition through publication credit. Among the disadvantages is the concern that if reproducing the artifact requires a “blow by blow” engagement with the author, that might means that it is not sufficiently well packaged and would require this engagement for any future reproduction (i.e., the artifact is not practically reproducible and reviewers end up debugging code), significantly higher time commitment for both authors and reviewers, and potential reviewer bias.
    \item \textbf{A hybrid approach}, balancing the advantages of both would involve a structured, limited-round rebuttal/revision process. Defined rebuttal stages enforce artifact improvement but at the same time limit time commitment  by limiting interactions to avoid unlimited “debugging obligations.” Some conferences implement what is sometimes called the ``kick-the-tiers phase" which effectively implements the first phase of such process.
\end{itemize}

Organizations should explicitly choose and document their interaction model based on their specific goals, available resources, and community preferences.

\noindent
\textbf{Recommendation 19 (organizations, community)}: Provide incentives for creating long-lasting artifacts, potentially awarded after a longer period of time (see also \cite{guilloteau2024longevity}). For computational artifacts, reproducibility has a limited shelf life that depends on a number of factors including hardware availability and software ecosystem status -- but primarily on the artifact quality and maintainability as well as the support the author is willing to give to an artifact. First, it was agreed that factors affecting artifact longevity are an important quality consideration that requires further study, development of best practices (e.g., in terms of using sound artifact source control and versioning), as well as improvement in terms of support ecosystem tooling by the community. Secondly, since hardware replacement cycles still pose a challenge to longevity, this provides an additional argument for using open testbeds as they are  more likely to keep hardware around specifically for reproducibility purposes). Last but not least, it is a critical factor to practical reproducibility and should be recognized by new incentives; specifically, Guilloteau suggests creating a new category of badge “artifact longevious” that indicates an artifact is following best practices for artifact longevity \cite{Guilloteau2024}.

\noindent
\textbf{Recommendation 20 (organizations)}: The HPC community would benefit from a dedicated coordinating forum or organization (either within existing professional societies or as a new entity) specifically focused on reproducibility practices. This entity should facilitate knowledge sharing between AE initiatives, develop standardized guidelines and checklists, maintain reviewer networks, create digital libraries of artifacts, and work with professional organizations (i.e., ACM and IEEE) to propose and eventually establish consistent reproducibility practices across conferences. By centralizing these efforts, such an organization could significantly improve efficiency, reduce duplication of work across conferences, and ensure that hard-won insights about practical reproducibility are preserved and disseminated throughout the community.

\section{Acknowledgments}

The workshop and report were supported by the REPETO project \cite{repeto} (NSF award 2226406) and indirectly by the Chameleon project \cite{chameleon_project} (NSF award 2027170).
\newpage
\section{Appendices}

\subsection{Overview}

Our appendices build upon several existing reproducibility resources in computing, drawing from conference-specific guidelines, i.e., USENIX Security \cite{usenix2023guidelines} and ACM's badging standards \cite{acm2020badges}; practical guides such as Barowy et al. \cite{barowy2019howto} and Padhye \cite{padhye2019tips}; broader frameworks like the Dagstuhl Beginners Guide \cite{bajpai2019dagstuhl}; platform recommendations from Code Ocean \cite{cheifet2021promoting}; systems research communities \cite{sysartifacts2024}; the cTuning Foundation's reproducibility checklist \cite{ctuning2020guidelines}; the FAIR principles \cite{wilkinson2016fair}; and community standards like SIGPLAN guidelines \cite{sigplan2020guidelines}.

While these resources provide valuable foundations and a specific knowledge base for some communities, i.e., ML, our workshop discussions identified persistent challenges specific to the HPC domain that remain insufficiently addressed. These appendices centralize, distill, and add several distinct contributions:

\begin{itemize}
    \item \textbf{HPC-Specific Focus}: The checklists are explicitly tailored for the unique challenges of HPC experimentation and reproducibility \cite{sedghpour2024artifactevaluation,guilloteau2024longevity,ANTUNES2024100655}, including specialized hardware requirements and extreme-scale computations.
    \item \textbf{Complementary Organization}: Appendix A systematically categorizes \emph{what} content must be included in reproducible artifacts (hardware, software, data, and longevity considerations), while appendix B addresses the often overlooked question of \emph{how} experiments should be structured and communicated.
    \item \textbf{Practical Reproducibility Conditions}: We emphasize recommendations that balance the cost of reproducibility with the benefits, e.g., establishing clear ``stopping conditions" for successful reproduction, particularly important for performance experiments where exact numerical reproduction may be impossible.
\end{itemize}

In developing these guidelines, we've distilled insights from our workshop discussions into practical, actionable recommendations for the HPC community. These appendices represent a tailored framework complementing existing reproducibility resources while addressing the specific needs of HPC research.

\subsection{Contributing to the Checklists}
\label{contributions}

The checklists in appendix A and B will continue to evolve as reproducibility practices mature in the HPC community. We invite community contributions through the following process:
GitHub Repository: The checklists are hosted on GitHub, making them easily accessible for community review and contribution.

\begin{itemize}
    \item \textbf{Contribution Mechanism}: Researchers can propose additions or modifications through GitHub \cite{ChameleonReproducibility2024} pull requests or by opening issues on the repository.
    \item \textbf{Review Process}: Our committee will review all contributions within 4 weeks of submission, evaluating them for practical utility and broad applicability across HPC domains.
    \item \textbf{Updates}: Approved contributions will be merged into the main branch, with attribution to contributors.
\end{itemize}

This collaborative approach ensures the checklists remain current and representative of best practices across the diverse HPC reproducibility landscape.

\newpage

\hypertarget{appendixa}{
    \subsection{Appendix A: Experiment Packaging Checklist}
}

\emph{WHAT to include for a complete reproducible artifact}

\subsubsection*{Hardware}
\begin{itemize}
    \item Are detailed hardware resource requirements quantified and documented accurately (memory, disk space, network bandwidth, CPU specs, GPU specs, etc.)?
    \item Are clear instructions provided on how to access the required hardware (e.g., testbeds, cloud, proprietary access)?
    \item Is the minimum viable scale for reproduction of large-scale experiments specified?
    \item Are any specific hardware configurations (BIOS settings, power modes) that affect results documented?
    \item If specialized hardware is required, is there guidance for identifying functional equivalents or fallback configurations?
\end{itemize}

\subsubsection*{Software}
\begin{itemize}
    \item Is all necessary source code with implementation details included?
    \item Are all software dependencies documented with specific version numbers or, if possible, specific repository commit hashes (which are static)?
    \item Has compatibility between licenses of all dependencies been verified and documented?
    \item Are configuration files with parameters used in the experiment provided?
    \item Are operating system requirements and tested versions specified?
    \item Is the execution environment (container or VM image, workflow definitions) packaged when possible and appropriate (e.g., when the build process is not itself an integral part of the "reproducibility" or stopping condition for your experiment)?
    \item Is the process for obtaining necessary licenses for proprietary software documented?
    \item Is the exact version of the artifact used in the paper explicitly documented (e.g., Git commit hash, release version, DOI)?
    \item Are security considerations and assumptions documented, particularly for artifacts requiring privileged access?
\end{itemize}

\subsubsection*{Data}
\begin{itemize}
    \item Are all necessary input datasets or generators included?
    \item For probabilistic experiments, are multiple input seeds or trial configurations provided to verify statistical significance?
    \item Are data versions, sources, and access methods specified?
    \item Are download scripts and persistent identifiers for large datasets provided to ensure data availability?
    \item Are sample or reduced datasets included for quick testing?
    \item Are any data preprocessing steps required documented?
    \item Are tools or scripts provided to validate input data integrity before experiments begin?
\end{itemize}

\subsubsection*{Artifact Longevity}
\begin{itemize}
    \item Is the expected maintenance period for the artifact (a "best by" date) specified?
    \item Are persistent identifiers and repositories used for long-term access?
    \item Are dependencies that may affect longevity documented?
    \item Is contact information for future questions included?
\end{itemize}

\newpage

\hypertarget{appendixb}{
    \subsection{Appendix B: Experiment Style Checklist}
}

\emph{HOW to organize and present artifacts for easier reviewer use}

\subsubsection*{Structure}
\begin{itemize}
    \item Is the experiment organized into distinct phases (setup, execution, analysis)?
    \item Is a top-level guide through the complete workflow provided?
    \item Are components modularized to allow partial execution, updates, and testing?
    \item Is a logical directory structure with descriptive names created?
\end{itemize}

\subsubsection*{Validation}
\begin{itemize}
    \item Are verification steps included after each major phase?
    \item Are expected output examples provided for validation points?
    \item Are automated tests that verify correct execution included?
    \item Are known warnings that can be safely ignored documented?
    \item Is troubleshooting guidance for common issues offered?
\end{itemize}

\subsubsection*{Configuration}
\begin{itemize}
    \item Is there a single configuration file or mechanism that controls all experiment parameters?
    \item Are configuration files or environment variables used instead of hardcoding values (including logins and passwords for accessing APIs/services)?
    \item Are all changeable values (paths, hostnames, file locations) parameterized?
    \item Are reasonable defaults included for all configurable parameters?
    \item Is experiment logic separated from environment-specific settings?
    \item Are there sanity checks that confirm the experimental environment is properly configured?
\end{itemize}

\subsubsection*{Reproducibility Condition}
\begin{itemize}
    \item Is a clear definition provided for what constitutes successful reproduction? Are both exact reproducibility criteria (e.g., bit-for-bit output matching) and functional reproducibility criteria (e.g., statistically equivalent results) defined (if applicable)?
    \item Are acceptable thresholds for numerical result variations specified?
    \item Are quantitative metrics provided to evaluate reproduction success?
    \item Are known sources of non-determinism and their expected impact documented? Are statistical methods provided to evaluate reproducibility (confidence intervals, error bars)?
    \item Are there tools or scripts to automatically compare and validate outputs against expected results?
\end{itemize}

\subsubsection*{Time \& Resource Estimation}
\begin{itemize}
    \item Are overall runtime estimates for the complete experiment provided?
    \item Is there a breakdown of time estimates for each major phase? Do distinct experiment stages have clear indicators of progress (i.e., a progress bar or estimated time left)?
    \item Are points where significant waiting periods occur indicated?
    \item Are any background processes that may impact timing noted?
    \item Are resource scaling properties documented (e.g., how requirements change with input size)?
    \item Is there guidance on optimizing resource usage for different hardware environments?
\end{itemize}

\newpage

\hypertarget{appendixc}{
    \subsection{Appendix C: Workshop Agenda and Discussion Prompts}
}

\newpage
\printbibliography
\end{document}